\documentclass[a4paper,fleqn,usenatbib]{mnras}

\usepackage[T1]{fontenc}
\usepackage{ae,aecompl}

\usepackage{graphicx}	%
\usepackage{amsmath}	%
\usepackage{amssymb}	%

\title[The Milky Way's rotation curve with SFDM]{The Milky Way's rotation curve with superfluid dark matter}

\author[S. Hossenfelder et al.]{
S. Hossenfelder,$^{1}$
T. Mistele$^{1}$\thanks{E-mail: mistele@fias.uni-frankfurt.de}
\\
$^{1}$Frankfurt Institute for Advanced Studies, Ruth-Moufang-Str. 1, D-60438 Frankfurt am Main, Germany
}

\date{Accepted XXX. Received YYY; in original form ZZZ}

\pubyear{2020}

\begin{document}
\label{firstpage}
\pagerange{\pageref{firstpage}--\pageref{lastpage}}
\maketitle

\begin{abstract}
Recent studies have shown that dark matter with a superfluid phase in which phonons mediate a long-distance force
gives rise to the phenomenologically well-established regularities of Modified Newtonian Dynamics ({\sc MOND}). Superfluid
dark matter, therefore, has emerged as a promising explanation for astrophysical observations by combining the benefits
of both particle dark matter and {\sc MOND}, or its relativistic completions, respectively.
We here investigate whether superfluid dark matter can reproduce the observed Milky Way rotation curve for $ R < 25\,\rm{kpc}$
and are able to answer this question in the affirmative.
Our analysis demonstrates that superfluid dark matter fits the data well with parameters in reasonable ranges.
The most notable difference between superfluid dark matter and {\sc MOND} is that superfluid dark matter requires about
$ 20\% $ less total baryonic mass (with a suitable interpolation function).
The total baryonic mass is then $5.96 \cdot 10^{10}\,M_\odot$, of which $1.03\cdot10^{10}\,M_\odot$ are from the bulge, $3.95\cdot10^{10}\,M_\odot$ are from the stellar disk, and $0.98\cdot10^{10}\,M_\odot$ are from the gas disk.
Our analysis further allows us to estimate the radius of the Milky Way's superfluid core (concretely, the so-called {\sc NFW} and thermal radii) and the
total mass of dark matter in both the superfluid and the normal phase.
By varying the boundary conditions of the superfluid to give virial masses $M_{200}^{\rm{DM}}$ in the range $0.5-3.0\cdot10^{12}\,M_\odot$, we find that the {\sc NFW} radius $R_{\rm{NFW}}$ varies between $65\,\rm{kpc}$ and $73\,\rm{kpc}$, while the thermal radius $R_T$ varies between about $67\,\rm{kpc}$ and $105\,\rm{kpc}$.
This is the first such treatment of a non-spherically-symmetric system in superfluid dark matter.
\end{abstract}


\section{Introduction}
\label{sec:introduction}

A naive combination of General Relativity an the matter-content of the Standard Model of particle physics does not correctly describe a variety of observations, ranging from the cosmic microwave background to galaxy clusters, to individual galaxies. The common remedy for this mismatch between theory and data is to conjecture a new type of matter -- ``dark matter'' -- which is presumably made of particles that so-far evaded direct detection.

However, in the past two decades it has become increasingly clear that particle dark matter has its own problems, problems that are particularly obvious on galactic scales. A promising, alternative explanation for the puzzling astrophysical observations has been around since the mid 1980s. Its original formulation has become known as Modified Newtonian Dynamics ({\sc MOND}) \citep{Milgrom1983a,Milgrom1983c,Bekenstein1984}. The  idea of {\sc MOND} is simple: Instead of increasing the gravitational pull by adding a new type of invisible matter, increase the gravitational pull of the normal matter by altering the force. For details, the reader is referred to the excellent reviews \citet{Sanders2002} and \citet{Famaey2012}.

{\sc MOND} is not relativistically invariant and non-local in the same way that Newtonian gravity is non-local. {\sc MOND} therefore must be understood as an approximation that has to be completed to give a fully-relativistic theory of modified gravity. A variety of such relativistic completions of {\sc MOND} have been proposed in the literature \citep{Bekenstein2004,Milgrom2010,Zlosnik2007,Milgrom2009,Deffayet2011,Blanchet2015,Hossenfelder2017b}, but they face a common problem. While modifications of gravity are superior on galactic scales because of their parametric simplicity \citep{Lelli2017b}, on larger scales dark matter is the simpler explanation. This tension between the cosmological and galactic scales can be resolved with any theory that combines cold dark matter for the former case and a {\sc MOND}-limit for the latter.
A very recent approach in this general direction is RelMOND \citep{Skordis2020}.
Another strong contender for such a theory is superfluid dark matter (hereafter: {\sc SFDM}) \citep{Berezhiani2015}.

The idea that dark matter may have a superfluid phase is itself not new \citep{Sikivie2009,Noumi2014,Davidson2013,DeVega2014,Davidson2015,Guth2015,Aguirre2016,Dev2017,Eby2018,Sarkar2018}, but the type of superfluid dark matter we are concerned with here is novel because it generates a long-range force. This force stems from the exchange of phonons between (effective) particles of normal matter and it reproduces the force-law of {\sc MOND}. If the superfluid phase however breaks down -- because the pressure is too low or the temperature too high -- then the matter will behave like common-type particle dark matter. One then has to find a suitable type of particle for which a superfluid exists in galaxies, but not in intergalactic space or in the early universe.

It was shown in \cite{Berezhiani2015,Berezhiani2018} that one can combine the successes of {\sc MOND} on galactic scales with the successes of {\sc $\Lambda$CDM} on cosmological scales by chosing an axion-like particle with a mass in the range of $\approx 1 {\rm eV}$. Besides the mass of the particle, the free parameters of this model are the self-interaction strength of the new particle (which has to generate a superfluid) and the strength of its interaction with baryonic matter. We know from the observation of a gravitational wave event with an electromagnetic counterpart \citep{Abbott2017} that the coupling of the superfluid to photons must be very small and we will therefore here set it to zero.

This {\sc SFDM} is still young and how well it will fare for large-scale structure formation is not presently known. It has been shown however that superfluid dark matter can correctly reproduce rotation curves for a large variety of galaxies by the same mechanism that {\sc MOND} does. For the same reason this model can explain the Tully-Fisher relation. In  \cite{Hossenfelder2019} it was furthermore demonstrated the {\sc SFDM} has no difficulties reproducing both the strong gravitational lensing data and the kinematic measurements for stellar rotation.

The aim of the present work is to see whether {\sc SFDM} can give a reasonable Milky Way rotation curve and, more generally, to gain a qualitative understanding of {\sc SFDM} in axisymmetric situations.
The rotation curve of the Milky Way in an axisymmetric approximation was previously discussed in \citet{Lisanti2019b}.
Here, we add to this a detailed comparison between {\sc SFDM}, its idealized {\sc MOND} limit, and the Radial Acceleration Relation in {\sc MOND}.
We also, for the first time, estimate the superfluid core's size and the virial dark matter mass of an axisymmetric system in {\sc SFDM}.
A statistically rigorous analysis of the goodness of fit of SFDM versus modified gravity and standard cold dark matter for the Milky Way rotation curve is beyond the scope of the present work.
Given the qualitative agreement of the {\sc SFDM} rotation curve with data that we will find below, such a study in isolation would anyway not be very informative.
Instead, one would have to take into account all available astrophysical constraints on the model's parameters.
The work presented here shows how the Milky-Way analysis could be done as part of a global fit that takes into account the superfluid core's finite size.

This paper is organized as follows. In Section \ref{sec:theory} we briefly summarize the key properties of superfluid dark matter and recall its relation to {\sc MOND}. In Section \ref{sec:data} we explain what data we are using. In Section \ref{sec:method} we detail how we integrate the equations and match them to the data. Results are presented in section \ref{sec:results}.  After a short discussion in Section \ref{sec:disc} we conclude in Section \ref{sec:conc}.

\section{Superfluid Dark Matter}
\label{sec:theory}

Following the notation of  \cite{Berezhiani2015}, we describe the superfluid by a massive scalar field with phase $\theta$. In the condensed phase the field associated with the phase is to good approximation classical. In the non-relativistic limit, the equations for the condensate are then \citep{Berezhiani2018, Hossenfelder2018}:
\begin{subequations}
\begin{align}
 \label{eq:poisson}
 \Delta \left( - \frac{\hat{\mu}}{m} \right) = 4 \pi G \left( \rho_b + \rho_{\rm{SF}}\left(\hat{\mu}, (\vec{\nabla} \theta)^2\right) \right) \,, \\
 \label{eq:phonon}
 \vec{\nabla} \left( \frac{(\vec{\nabla} \theta)^2 + 2m (\frac{2 \beta}{3} -1) \hat{\mu} }{\sqrt{(\vec{\nabla} \theta)^2 + 2m(\beta-1)\hat{\mu}}}  \vec{\nabla} \theta \right) =  \frac{\alpha}{2 M_{\rm{Pl}}} \, \rho_b \,,
\end{align}
\end{subequations}
where
\begin{align}
 \rho_{\rm{SF}}\left(\hat{\mu}, (\vec{\nabla} \theta)^2\right) = \frac{2\sqrt{2}}{3} m^{5/2} \Lambda \, \frac{3(\beta-1) \hat{\mu} + (3-\beta) \frac{(\vec{\nabla} \theta)^2}{2m}}{\sqrt{ (\beta-1) \hat{\mu}  + \frac{(\vec{\nabla} \theta)^2}{2m} }} \,.
\end{align}

Here,  $ \rho_{\rm{SF}} $ is the energy-density of the superfluid and $ \rho_b $ is the energy-density of baryons whose profile we extract from data as described in Sec.~\ref{sec:data}. $m$ is the mass of the superfluid's constituent particles, and $ \hat{\mu} = \mu_{\rm{nr}} - m \, \phi_{\rm{N}} $ is a combination of the non-relativistic chemical potential $ \mu_{\rm{nr}}$ (a constant that acts as initial value) and the Newtonian gravitational potential $ \phi_{\rm{N}} $. Expressed in these terms,
Eq.~\eqref{eq:poisson} is the usual Poisson equation for the Newtonian gravitational potential and
Eq.~\eqref{eq:phonon} determines {\sc SFDM}'s phonon field $ \theta $ that carries the {\sc MOND}-like force.

In the above equations, $ M_{\rm{Pl}} $ denotes the Planck mass, $ \alpha $ is a dimensionless coupling constant, and $ \Lambda $ is related to the self-interaction strength of the new field \citep{Berezhiani2015}.
The parameter $ \beta $ quantifies finite-temperature corrections as discussed in \cite{Berezhiani2015}.

In {\sc SFDM}, the baryons' total acceleration $\vec a_{\rm tot}$ is then a sum of the acceleration from the Newtonian gravitational pull due to the baryonic and superfluid mass densities, $\vec a_{\rm N}$, and the acceleration from the phonon force, $\vec a_\theta$, which is proportional to the gradient $ \vec{\nabla} \theta $:
\begin{align}
 \vec{a}_{\rm{tot}} = \vec{a}_{\rm{N}} + \vec{a}_\theta = -\vec{\nabla} \phi_{\rm{N}} - \frac{\alpha \Lambda}{M_{\rm{Pl}}} \vec{\nabla} \theta \,.
\end{align}

In our below analysis, we will further consider an idealized {\sc MOND}-limit to highlight the differences between the two models. In this {\sc MOND}-limit, we neglect the energy-density of the superfluid and assume that the kinetic energy of the phonon field is much larger than the chemical potential $ |\vec{\nabla} \theta|^2/(2m) \gg \hat{\mu} $.
In this case, the equations simplify to
\begin{subequations}
\begin{align}
 \label{eq:poissonmond}
 \Delta \phi_{\rm{N}} &= 4 \pi G \, \rho_b \,, \\
 \label{eq:mond}
 \vec{\nabla} \left( \left(\frac{\sqrt{(\vec{\nabla} \bar{\theta})^2}}{a_{0,\theta}} \right) \vec{\nabla} \bar{\theta} \right) &=  4 \pi G \, \rho_b \,,
\end{align}
\end{subequations}
where we introduced the parameters $ \bar{\theta} = (\alpha \Lambda/M_{\rm{Pl}}) \theta $, $ a_{0,\theta} = \alpha^3 \Lambda^2/M_{\rm{Pl}} $, and  $ 8 \pi G = M_{\rm{Pl}}^{-2} $ to match the expressions to the common {\sc MOND}-formalism.

{\sc MOND} further requires an interpolation function whose purpose is to fade out the regime of Newtonian gravity and cross over to the new, logarithmic potential law. This interpolation function is defined as the scalar function that, when multiplied with the acceleration from the Newtonian gravitational pull of the normal matter gives the total acceleration in {\sc MOND}. This interpolation function corresponds to the phenomenologically obtained Radial Acceleration Relation ({\sc RAR})\footnote{We wish to emphasize that with Radial Acceleration Relation we refer to the relation between the radial accelerations in general, not to the fit of this relation with a specific interpolation function.}. Strictly speaking, this is correct only in the no-curl approximation, see Section \ref{sec:method}. However, we show in Section \ref{sec:mond} that the no-curl approximation is good to use, and hence the use of the radial acceleration relation justified.

An interpolation function is not necessary in {\sc SFDM}, but to compare our results to those obtained with {\sc MOND}, we use the exponential {\sc MOND} interpolation function
\begin{align}
 \label{eq:nuRAR}
 \nu_{\rm{e}}(y) = \frac{1}{1 - e^{-\sqrt{y}}} \,,
\end{align}
where $ y = a_{\rm{b}}/a_{0,{\rm{e}}} $ with the Newtonian baryonic acceleration $ a_{\rm{b}} $ and the acceleration scale $ a_{0,\rm{e}} = 1.2\cdot10^{-10}\,\rm{m}/\rm{s}^2 $.

As laid out in \cite{Hossenfelder2018}, in the idealized {\sc MOND}-limit, {\sc SFDM} is equivalent to {\sc MOND} with the interpolation function
\begin{align}
 \label{eq:nuSFDM}
 \nu_{\theta}(y) = 1 + \frac{1}{\sqrt{y}} \,,
\end{align}
where $ y = a_{\rm{b}}/a_{0,\theta} $ and the acceleration scale is $ a_{0,\theta} = 0.87\cdot10^{-10}\,\rm{m}/\rm{s}^2 $ for the fiducial parameter values from \citet{Berezhiani2018}. Again, this equivalence strictly speaking only holds in the no-curl-approximation, because the curl-terms in the {\sc SFDM} model are different from those of {\sc MOND} even in the idealized {\sc MOND}-limit. However, as we will see later, the difference is negligible for fitting the data we will be dealing with.
In the non-idealized case, {\sc SFDM} cannot be cast into the form of an interpolation function both because of the superfluid's gravitational pull and because the phonon force is not exactly of the {\sc MOND} form.

\section{Data}
\label{sec:data}

Recently, \cite{McGaugh2019b} put forward a new model for the Milky Way ({\sc MW}) to match the most up-to-date terminal velocity data. It provides an excellent fit to the {\sc MW}'s rotation curve even outside the range where this model was fitted. This model relates the Newtonian acceleration from the baryonic mass distribution with the total acceleration by employing the Radial Acceleration Relation from \cite{Lelli2017b}.

We take the baryonic mass distribution from \cite{McGaugh2019b} with a few minor modifications.
Our model consists of a bulge, a gas disk, and a stellar disk.
We take the stellar disk exactly as in \cite{McGaugh2019b}, i.e. we use a scale height of $ 300\,\rm{pc} $ with the numerical surface density from \cite{McGaugh2019b}.
For the gas disk, we also use the numerical surface density from \cite{McGaugh2019b}.
However, following \cite{Bovy2013}, we use a disk with scale height $ 130\,\rm{pc} $ instead of an infinitely thin disk
because a finite height is easier to handle with our numerical code.
The choice of this scale height does not significantly affect our results.

The gas surface density is that of \cite{Olling2001}, but scaled up by a factor of $ 1.4 $ to account for helium and metals, and adjusted for newer measurements of the Galactic size $ R_0 $ \citep{GravityCollaboration2018}, see \cite{McGaugh2008, McGaugh2019b} for details.
This value of $R_0$ is a little outdated and a newer value was provided in \citet{GravityCollaboration2019}. However, since the exact value does not affect our conclusions much we just reuse the model from \cite{McGaugh2019b} with the therein used value.
The resulting gas surface density in this model is not a smooth exponential.
The stellar surface density is based on a Freeman Type II profile \citep{Freeman1970} that was adjusted to fit the detailed terminal velocity data for $ 3\,\rm{kpc} < R < 8\,\rm{kpc} $, see \cite{McGaugh2019b}.

The bulge profile is parameterized as \cite{McGaugh2008}
\begin{align}
 \label{eq:bulge}
 \rho_{\rm{bulge}}(b) = \frac{\rho_{\rm{bulge},0}}{\eta \zeta b_m^3} \frac{\exp\left[ -(b/b_m)^2\right]}{\left(1+ b/b_0\right)^{1.8}}~,
\end{align}
where $ \eta = 0.5 $, $ \zeta = 0.6 $, $b_m = 1.9\,\rm{kpc} $, $ b_0 = 0.1\,\rm{kpc} $, and $ b = r/(\eta \zeta)^{1/3} $ with the spherical radius $ r $.
This is the spherically equivalent mass distribution of the triaxial model used in \cite{McGaugh2019b}.
The constant $ \rho_{\rm{bulge},0} $ is chosen such that the asymptotic Newtonian acceleration due to the bulge is the same as in \cite{McGaugh2019b}, see Table~2 there.

In the following, we keep the shape of the baryonic mass distribution fixed, but allow to rescale the baryonic mass distribution as a whole in order to fit the {\sc MW} rotation curve.
For this, we introduce the parameter $ f_b $ that multiplies the baryonic mass distribution $ \rho_b(R, z) $.
Here, $ R $ and $ z $ are the usual cylindrical coordinates.

We take the {\sc MW} rotation curve data from \cite{McGaugh2019b}.
That is, we take the rotation curve data from \cite{Eilers2019} for $ R > 5\,\rm{kpc} $ and from \cite{Portail2017} for $ R < 2.2\,\rm{kpc} $, but with two adjustments made in \cite{McGaugh2019b}:
The Jeans analysis in \cite{Eilers2019} assumes a smooth exponential stellar profile.
Therefore, \cite{McGaugh2019b} has redone the analysis of \cite{Eilers2019} with the profile described above.
Also, \cite{McGaugh2019b} has scaled the radii of \cite{Portail2017} to be consistent with the newer measurement of $ R_0 $ from \cite{GravityCollaboration2018}.
Rescaling the total baryonic mass with our parameter $ f_b $ does not require redoing the analysis from \cite{Eilers2019}, since the normalization cancels out, see e.g. their Eq.~(3).
Therefore, we can adjust the total baryonic mass while keeping the same rotation curve data.

\section{Method and Data Analysis}
\label{sec:method}

\subsection{The Rotation Curve}

Equations \eqref{eq:poisson} and \eqref{eq:phonon} describe only the core of the superfluid in the galactic center, not the non-superfluid phase at larger radii. However, our later estimate (see Section \ref{sec:sfdm:sfcore}) shows that all data points fall well inside the superfluid core so
that Eqs.~\eqref{eq:poisson} and \eqref{eq:phonon} are sufficient to calculate the {\sc MW} rotation curve in {\sc SFDM}. As numerical values for the parameters in Eqs.~(\ref{eq:poisson}) and (\ref{eq:phonon}) we use the fiducial values from \cite{Berezhiani2018}: $ \beta = 2 $, $ \alpha = 5.7 $, $ m = 1\,\rm{eV} $, and $ \Lambda = 0.05 \, \rm{meV} $.

To integrate the equations, we assume axisymmetry and, as usual, parameterize it with cylindrical coordinates $ (R, z) $. We calculate the rotation curve $ v_c(R) $ as
\begin{align} \label{vc}
 v_c(R) = \sqrt{R \cdot |a_{\rm{tot}, R}(R,z=0)|}\,,
\end{align}
where $ a_{\rm{tot}, R} $ is the $R$-component of the total acceleration $ \vec{a}_{\rm{tot}} $.
 In the case of the full {\sc SFDM} equations we use the boundary conditions
\begin{subequations}
\begin{align}
 \left. \partial_z \hat{\mu} \right|_{z=0} &= 0 \,, \\
 \left. \partial_z \theta \right|_{z=0} &= 0 \,, \\
 \left. \hat{\mu} \right|_{\sqrt{R^2+z^2} = r_\infty} &= \mu_\infty \,, \\
 \left. \theta \right|_{\sqrt{R^2+z^2} = r_\infty} &= 0 \,.
\end{align}
\end{subequations}
The first two conditions encode the $ z \to -z $ symmetry, the other two conditions impose spherical symmetry at $ r_\infty $.
This approximation of spherical symmetry is a good approximation in {\sc MOND} and Newtonian gravity \citep{Milgrom1986} which makes it reasonable
that it is a good approximation here too.

For $ \theta $, the numerical value at $ r_\infty $ does not enter our equations so we just set it to zero.
For $ \hat{\mu} $, the numerical value at $ r_\infty $ determines the superfluid's density and therefore the superfluid's gravitational pull.
This gravitational pull is typically subdominant in the inner regions of a galaxy, but it is important at larger radii, e.g. to produce enough strong lensing \citep{Hossenfelder2018}.
Here, we choose the fixed value $ \mu_\infty/m = 1.25\cdot10^{-8} $ at $r_\infty = 100\,\rm{kpc}$.
As we will see below, this gives a subdominant but non-negligible contribution to the rotation curve at $ R < 25\,\rm{kpc} $.
Other choices of $\mu_\infty/m$ and $r_\infty$ are discussed in Sec.~\ref{sec:total}.

For the idealized {\sc MOND} limit, we impose the boundary conditions
\begin{subequations}
\begin{align}
 \left. \partial_z \phi_{\rm{N}} \right|_{z=0} &= 0 \,, \\
 \left. \partial_z \bar{\theta} \right|_{z=0} &= 0 \,, \\
 \left. \phi_{\rm{N}} \right|_{\sqrt{R^2+z^2} = r_\infty} &= 0 \,, \\
 \left. \bar{\theta} \right|_{\sqrt{R^2+z^2} = r_\infty} &= 0 \,.
\end{align}
\end{subequations}
Since in this limit we neglect the superfluid's energy density $ \rho_{\rm{SF}} $, the numerical value of $ \phi_{\rm{N}} $ and $ \bar{\theta} $ do not enter the equations and we can set both to zero.

Eqs.~\eqref{eq:mond} and \eqref{eq:phonon} are of the form $ \vec{\nabla} ( g \, \vec{\nabla} \theta ) = 4 \pi G \, \rho_b $ with some function $ g $.
Therefore, they can be written as $ \vec{\nabla} ( g \, \vec{\nabla} \theta - \vec{\nabla} \phi_{\rm{N},b} ) = 0 $, where $ \phi_{\rm{N},b} $ is the Newtonian gravitational potential produced by the baryons.
This gives $ g \, \vec{\nabla} \theta = \vec{\nabla} \phi_{\rm{N},b} $ up to a term that can be written as the curl of a vector field.
Neglecting this term is often a reasonable approximation \citep{Brada1995}.
Below, we will refer to this approximation as the `no-curl-approximation' and investigate how good an approximation to the full equations it is. 

A summary of the numerical parameters used in the present work is shown in Table~\ref{tab:params}. For more details on our numerical analysis, please refer to Appendix~\ref{sec:numerical}.

\begin{table}
\center
\caption{The numerical parameters used in the present work. We keep the model parameters fixed at the fiducial values from \citet{Berezhiani2018}. The baryonic density is fixed up to an overall factor $f_b$, various values of which are discussed in Sec.~\ref{sec:mond} and Sec.~\ref{sec:sfdm}. The boundary conditions for the superfluid are kept fixed in Sec.~\ref{sec:mond} and Sec.~\ref{sec:sfdm} but are varied in Sec.~\ref{sec:total}.}
\label{tab:params}
\begin{tabular}{l|c|c|}
\hline
\multicolumn{3}{|l|}{Model parameters} \\
\hline
$m$ & $1\,\rm{eV}$ & from \citet{Berezhiani2018} \\
$\alpha$ & $5.7$ & from \citet{Berezhiani2018} \\
$\Lambda$ & $0.05\,\rm{meV}$ & from \citet{Berezhiani2018} \\
$\beta$ & $2$ & from \citet{Berezhiani2018} \\
\hline
\multicolumn{3}{|l|}{Baryonic mass} \\
\hline
Bulge & $1.29\cdot10^{10}\,M_\odot \times f_b$ & see Eq.~\eqref{eq:bulge} \\
Stellar disk & $4.94 \cdot 10^{10}\,M_\odot \times f_b$ & from \citet{McGaugh2019b} \\
Gas disk & $1.22 \cdot 10^{10}\,M_\odot \times f_b$ & from \citet{McGaugh2019b} \\
$f_b$ & $0.8$ & unless stated otherwise \\
\hline
\multicolumn{3}{|l|}{Boundary condition} \\
\hline
$\mu_\infty/m$ & $1.25\cdot10^{-8}$ & unless stated otherwise \\
$r_\infty$ & $100\,\rm{kpc}$ & unless stated otherwise \\
\hline
\end{tabular}
\end{table}

\subsection{The Size of the Superfluid Core}
\label{sec:sfcoremethods}

\cite{Berezhiani2018} gives two different methods to estimate the size of the superfluid core for the case of spherical symmetry.
The first method uses the so-called thermal radius $ R_{\rm{T}} $, the second method uses the so-called {\sc NFW} radius $ R_{\rm{NFW}} $.
Here, we will generalize both methods to axisymmetric situations.
This will allow us to estimate the size of the superfluid's core in the $R$-direction ($ R_{\rm{T},R} $ and $ R_{\rm{NFW},R} $) and in the $z$-direction ($ R_{\rm{T},z} $ and $ R_{\rm{NFW},z} $).
Both methods will make use of the fact that the superfluid core is approximately spherically symmetric at large radii, although it is only axially symmetric at smaller radii.

We start with the thermal radius $ R_{\rm{T}} $.
According to Sec.~III of \cite{Berezhiani2018} this radius is determined by the relation
\begin{align}
 \label{eq:RT}
 \Gamma = t_{\rm{dyn}}^{-1}\,,
\end{align}
where $ \Gamma $ is the local self-interaction rate and $ t_{\rm{dyn}} $ is the dynamical time.
Here, $ \Gamma = (\sigma / m) \, \mathcal{N} \, v \, \rho $, where $\sigma$ is the self-interaction rate, $ \mathcal{N} = (\rho/m) (2\pi/mv)^3 $
is the Bose-degeneracy factor, and $v$ is the average velocity of the particles.
As in \cite{Berezhiani2018}, we take $ \sigma/m = 0.01\,\rm{cm}^2/\rm{g} $.

The assumption of spherical symmetry enters in the calculation of $ t_{\rm{dyn}} $ and $ v $.
Specifically, \cite{Berezhiani2018} takes $ t_{\rm{dyn}} \approx r/v $ and $ v^2 \approx r \partial_r \phi_{\rm{N}} $ with the spherical radius $ r $.
For axisymmetric situations, we adjust this to be $ t_{\rm{dyn}} \approx R/v|_{z=0} $ and $ v^2 \approx R \partial_R \phi_{\rm{N}}|_{z=0} $ for the thermal radius $ R_{\rm{T},R} $ in $R$-direction and $ t_{\rm{dyn}} \approx z/v|_{R=0} $ and $ v^2 \approx z \partial_z \phi_{\rm{N}}|_{R=0} $ for the thermal radius $ R_{\rm{T},z} $ in $ z$-direction.

To estimate the {\sc NFW} radius $ R_{\rm{NFW}} $, one assumes a superfluid in the centers of galaxies followed by an {\sc NFW} profile at larger radii.
The {\sc NFW} radius $ R_{\rm{NFW}} $ is then the radius at which the density and pressure of the superfluid core can be matched to the density and pressure of an {\sc NFW} halo.
Since the standard {\sc NFW} profile is by definition spherically symmetric, the {\sc NFW} radius is well-defined only when the superfluid core is spherically symmetric as well at the radius $ R_{\rm{NFW}} $.
In our case, the superfluid will not be exactly spherically symmetric.
However, the superfluid is approximately spherically symmetric at the {\sc NFW} radius, if this radius is large enough.
Thus, we define $ R_{\rm{NFW}, R} $ to be the value of $ R $ at $ z = 0 $, where the superfluid's density and pressure match those of the {\sc NFW} halo, and $ R_{\rm{NFW},z} $ to be the value of $ z $ at $ R = 0 $ where the pressure and density match.
If the superfluid is approximately spherically symmetric at the {\sc NFW} radii defined in this way, we will have $ R_{\rm{NFW},R} \approx R_{\rm{NFW},z} $.
The formulas for the pressure and density of the superfluid and the {\sc NFW} halo can be taken directly from \cite{Berezhiani2018}.

\section{Results}
\label{sec:results}

\subsection{Results for the idealized {\sc MOND} regime}
\label{sec:mond}

We begin with the results for the idealized {\sc MOND}-limit of superfluid dark matter.  The free parameter in this fit, $f_b$, is a factor for rescaling the total mass of baryons. In \cite{McMillan2017, Licquia2015} the stellar mass in the Milky Way was estimated with an observational uncertainty in the range of $10-20\%$. We therefore expect that the total mass of baryons has a similar observational uncertainty, which justifies allowing $f_b$ to vary by this amount.

\begin{figure}
 \includegraphics[width=\columnwidth]{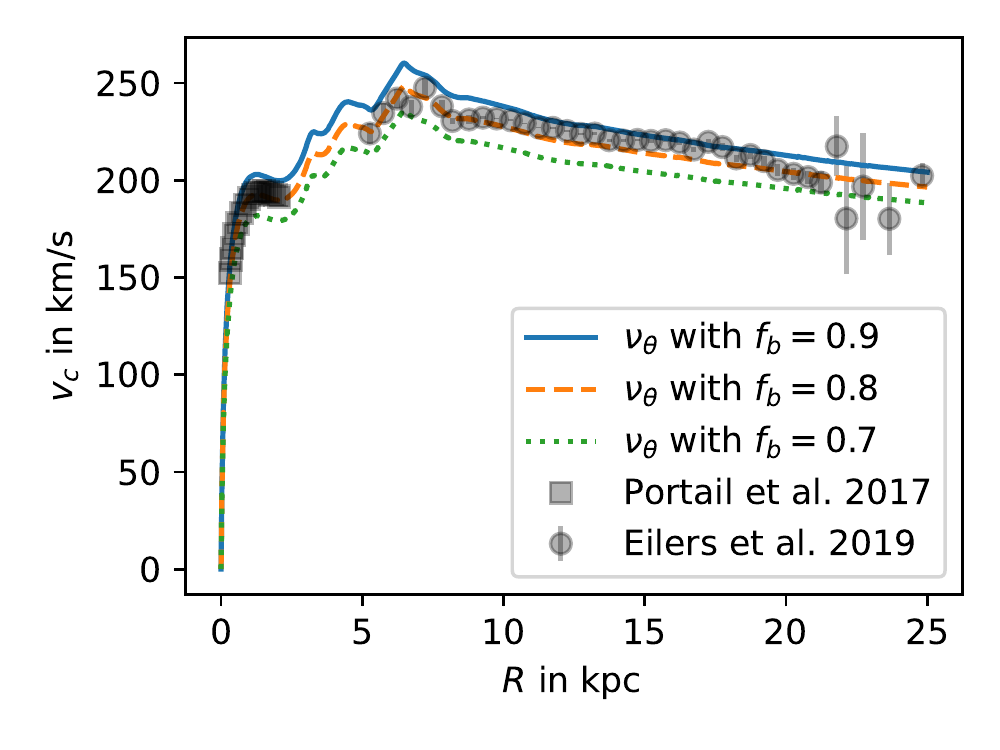}
 \caption{Milky Way rotation curve in the idealized {\sc MOND}-limit of {\sc SFDM} for different values of $f_b$.
 Also shown are the data from \citet{Eilers2019} (black dots) and \citet{Portail2017} (black squares), both adjusted to match the assumptions of \citet{McGaugh2019b}.}
 \label{fig:MONDSFDMandRARrot}
\end{figure}

Fig.~\ref{fig:MONDSFDMandRARrot}
shows the rotation curve of the Milky Way in terms of $v_c$ from Eq. (\ref{vc}) in the {\sc MOND}-limit in comparison with data for $ f_b = 0.9, 0.8 $, and $ 0.7 $. The rotation curve data in this figure is that from \cite{Eilers2019} and \cite{Portail2017} adjusted to match the assumptions of \cite{McGaugh2019b}, as described in Sec.~\ref{sec:data}.

As one sees, to obtain a reasonable fit of the rotation curve data in the idealized {\sc MOND}-limit of {\sc SFDM}, we need $ 10-20\,\% $ less baryonic mass than the model of \cite{McGaugh2019b}, i.e. $ f_b \approx 0.8-0.9 $.
For $ f_b = 0.8 $, the total stellar mass in our model is $ M_* = 4.98 \cdot 10^{10}\,M_\odot $.
\cite{McMillan2017} estimates the {\sc MW}'s total stellar mass as $(5.43\pm 0.57) \cdot 10^{10}\,M_\odot $ and \cite{Licquia2015} estimates $(6.08\pm 1.14)\cdot 10^{10}\,M_\odot $.
Thus, our value for $ M_* $ is relatively low, but still falls inside the error bars of \cite{McMillan2017, Licquia2015}. 

We have checked that with the $ \nu_{\rm{e}} $ interpolation function and the no-curl-approximation for $ f_b = 1 $, our calculation reproduces  the results shows in Fig.~3 of \cite{McGaugh2019b} up to numerical differences and the minor modifications described in Sec.~\ref{sec:data}.

There are three differences between {\sc SFDM}'s idealized {\sc MOND} regime and the {\sc RAR} as applied in \cite{McGaugh2019b}.
First, the shapes of the interpolation functions $ \nu_\theta $ and $ \nu_{\rm{e}} $ differ.
Second, the acceleration scales $ a_{0,\theta} $ and $ a_{0,\rm{e}} $ differ.
And third, \cite{McGaugh2019b} uses the no-curl-approximation while {\sc SFDM}'s idealized {\sc MOND}-regime does not.

We have found that using the no-curl-approximation induces a non-negligible error only at $ R \lesssim 5\,\rm{kpc} $.
But even at $ R \lesssim 5\,\rm{kpc} $, this error is only a few percent on  $ v_c $.
Thus, the no-curl-approximation is not the main reason {\sc SFDM}'s idealized {\sc MOND} regime requires significantly less baryonic mass than the $ \nu_{\rm{e}} $-based model from \cite{McGaugh2019b}.

\begin{figure}
 \includegraphics[width=\columnwidth]{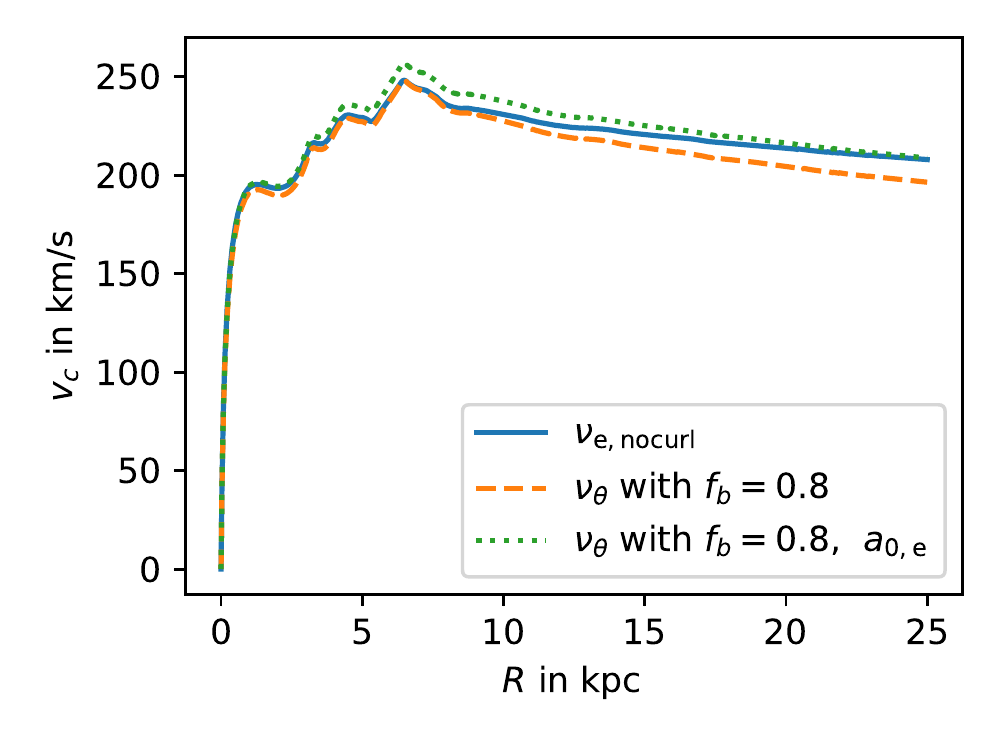}
 \caption{Top: Milky Way rotation curve in the idealized {\sc MOND} regime of {\sc SFDM} for $ f_b = 0.8 $ (dashed orange line), with $ f_b = 0.8 $ and $ a_{0,\rm{e}} $ instead of $ a_{0,\theta} $ (dotted green line) compared with the $ \nu_{\rm{e}} $ rotation curve with $ f_b = 1 $ (solid blue line).
}
 \label{fig:MONDSFDMshapeAndCurl}
\end{figure}

The effect of the different acceleration scales $ a_{0,\theta} $ and $ a_{0,\rm{e}} $ can be seen by using $ a_{0,\rm{e}} $ instead of $ a_{0,\theta} $ in the equations of {\sc SFDM}'s idealized {\sc MOND} regime.
This is shown in Fig.~\ref{fig:MONDSFDMshapeAndCurl}.
The effect of using  $ a_{0,\rm{e}} $ instead of $ a_{0,\theta} $ is larger at larger radii, where the additional {\sc MOND}-like force begins to dominate.
There, the larger acceleration scale increases $ v_c $.
Thus, the smaller acceleration scale $ a_{0,\theta} $ helps {\sc SFDM}'s idealized {\sc MOND}-limit to not require even less mass to fit the {\sc MW} rotation curve.

This leaves the shape of $ \nu_\theta $ as the main reason that {\sc SFDM}'s idealized {\sc MOND} regime requires less baryonic mass than the model from \cite{McGaugh2019b}.
Concretely, $ \nu_{\rm{e}} $ approaches $ 1 $ at large baryonic accelerations much faster than $ \nu_\theta $.
Therefore, at large and intermediate accelerations, $ \nu_\theta $ produces a larger total acceleration than $ \nu_{\rm{e}} $.
As a result, less baryonic mass is needed to match the rotation curve data.

From Fig.~\ref{fig:MONDSFDMshapeAndCurl} we can further see that {\sc SFDM}'s idealized {\sc MOND} regime produces rotation curves not only with a different normalization compared to the model from \cite{McGaugh2019b}, but also with a different shape.
This is a consequence of both the different interpolation functions $ \nu_{\rm{e}} $ and $ \nu_\theta $ and the different acceleration scales $ a_{0,\theta} $ and $ a_{0,\rm{e}} $.
Below, we will see that the full {\sc SFDM} equations can produce rotation curves that are closer to that of the $ \nu_{\rm{e}} $-based model from \cite{McGaugh2019b}.

The above results for {\sc SFDM}'s idealized {\sc MOND}-limit may also apply to Covariant Emergent Gravity ({\sc CEG}) \citep{Hossenfelder2017b} because {\sc CEG} reduces to the same equations as {\sc SFDM}'s idealized {\sc MOND}-limit, if we assume that only the 0-component of {\sc CEG}'s vector field $ u^\mu $ is nonzero \citep{Hossenfelder2018}.
However, it is not clear whether or not this assumption holds for axisymmetric systems like the {\sc MW}.
Also, the numerical value of the acceleration scale may be different for {\sc CEG}, see \cite{Hossenfelder2018}.
Thus, one should be careful when applying the above results to {\sc CEG}.

\subsection{Results for full {\sc SFDM}}
\label{sec:sfdm}

\begin{figure}
 \includegraphics[width=\columnwidth]{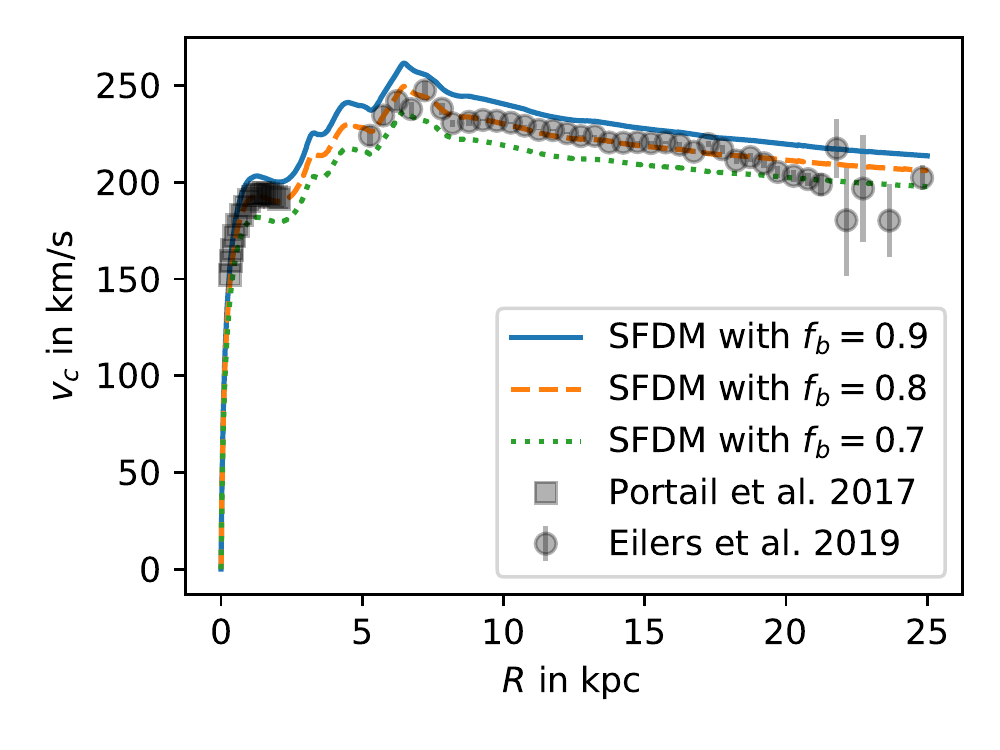}
 \caption{{\sc SFDM} {\sc MW} rotation curve for $ f_b = 0.9 $ (blue line), $ f_b = 0.8 $ (dashed orange line), and $ f_b = 0.7 $ (dotted green line).
 Also shown are the data from \citet{Eilers2019} (black dots) and \citet{Portail2017} (black squares), both adjusted to match the assumptions of \citet{McGaugh2019b}.}
 \label{fig:SFDMrot}
\end{figure}

Fig.~\ref{fig:SFDMrot} compares the data of the Milky-Way rotation curve to the results from the full equations of superfluid dark matter.
Again we have plotted the results for different values of the total mass of baryons, $ f_b = 0.7, 0.8 $, and $ 0.9 $. As one sees, we get a good fit for $ f_b $ around $ 0.8 $, which is similar to our finding for the idealized {\sc MOND}-limit discussed in the previous subsection.

\begin{figure}
 \includegraphics[width=\columnwidth]{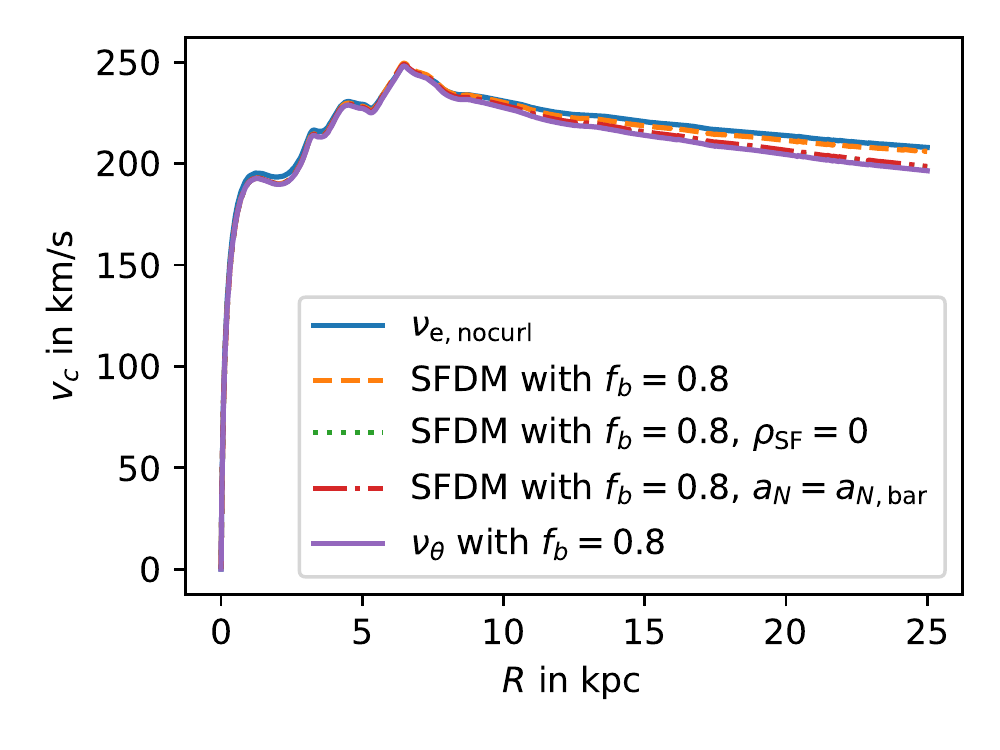}
 \caption{
     {\sc MW} rotation curve from the $ \nu_{\rm{e}} $-based model from \citet{McGaugh2019b} (solid blue line), from {\sc SFDM} with $ f_b = 0.8 $ (dashed orange line), from {\sc SFDM}'s idealized {\sc MOND} limit with $ f_b = 0. 8 $ (solid purple line), and from {\sc SFDM} with $f_b = 0.8 $ with the superfluid's gravitational pull removed using two different methods.
     The first method is to set $ \rho_{\rm{SF}} = 0 $ in {\sc SFDM}'s equations (dotted green line).
     The second method is to keep $ \rho_{\rm{SF}} $, but use only the baryonic Newtonian acceleration when calculating the rotation curve (dash-dotted red line).
}
 \label{fig:SFDMrothalo}
\end{figure}

This is not a trivial consequence of the results from the realized {\sc MOND} limit because now the superfluid's gravitational pull also contributes to the rotation curve.
This gravitational pull can change both the shape and the normalization of the rotation curve.
In our case, the superfluid's gravitational pull has a significant effect on the shape of the rotation curve.
Indeed, the shape of the rotation curve is now closer to the rotation curve obtained using the $ \nu_{\rm{e}} $ interpolation function in \cite{McGaugh2019b}.
This becomes clear from Fig.~\ref{fig:SFDMrothalo} which compares the rotation curve for {\sc SFDM} with $ f_b = 0.8 $ (dashed orange line), the rotation curve obtained using the $ \nu_{\rm{e}} $ interpolation function with $ f_b = 1 $ (solid  blue line), and the rotation curve obtained in {\sc SFDM}'s idealized {\sc MOND}-limit for $ f_b = 0.8 $ (solid purple line).

This change of shape is due to the superfluid's gravitational pull as one can see by removing the superfluid's gravitational pull from the full {\sc SFDM} model.
The resulting rotation curves are very close to the rotation curve of {\sc SFDM}'s idealized {\sc MOND}-limit, as shown in Fig.~\ref{fig:SFDMrothalo}.
For this, we have tried two different methods of removing the superfluid's gravitational pull from the full {\sc SFDM} model, but both give similar results.

The first method is to simply set $ \rho_{\rm{SF}} = 0 $.
However, this does not just remove the superfluid's gravitational pull but also influences the $ \theta $ equation of motion, i.e. the equation that determines the phonon force.
This is because $ \hat{\mu} $ enters the $ \theta $ equation of motion, but $ \hat{\mu} $ is different with $ \rho_{\rm{SF}} = 0 $ and $ \rho_{\rm{SF}} \neq 0 $.

This motivates the second method of removing the superfluid's gravitational pull for which we solve the full {\sc SFDM}-equations without removing $ \rho_{\rm{SF}} $, but when calculating the rotation curve we take $ \vec{\nabla} \phi_{\rm{N}} $ to be only the baryonic Newtonian gravitational pull, not the full Newtonian gravitational force including $ \rho_{\rm{SF}} $.
This method is more ad-hoc, but has the advantage that the solution for $ \theta $ is not affected.

However, as one sees in Fig.~\ref{fig:SFDMrothalo}, the difference between both methods is small.
Both methods give a rotation curve that is very close to that of the idealized {\sc MOND}-limit.
Therefore, we can unambiguously attribute the shape difference between the full model and the idealized {\sc MOND}-limit to the superfluid's gravitational pull.

\subsection{The Size of the Superfluid Core}
\label{sec:sfdm:sfcore}

In {\sc SFDM}, galaxies contain a superfluid phase only at the centers of galaxies.
In the outer parts of galaxies, the superfluid breaks down.
The equations used above are valid only in the superfluid phase.
Therefore, our results regarding the {\sc MW}'s rotation curve are valid only if the superfluid phase extends to $ R > 25\,\rm{kpc} $.
To check that it is consistent to use only the superfluid phase to fit the rotation curve data, we therefore estimated the size of the {\sc MW}'s superfluid core using
the methods outlined in \ref{sec:sfcoremethods}.

\begin{figure}
 \includegraphics[width=\columnwidth]{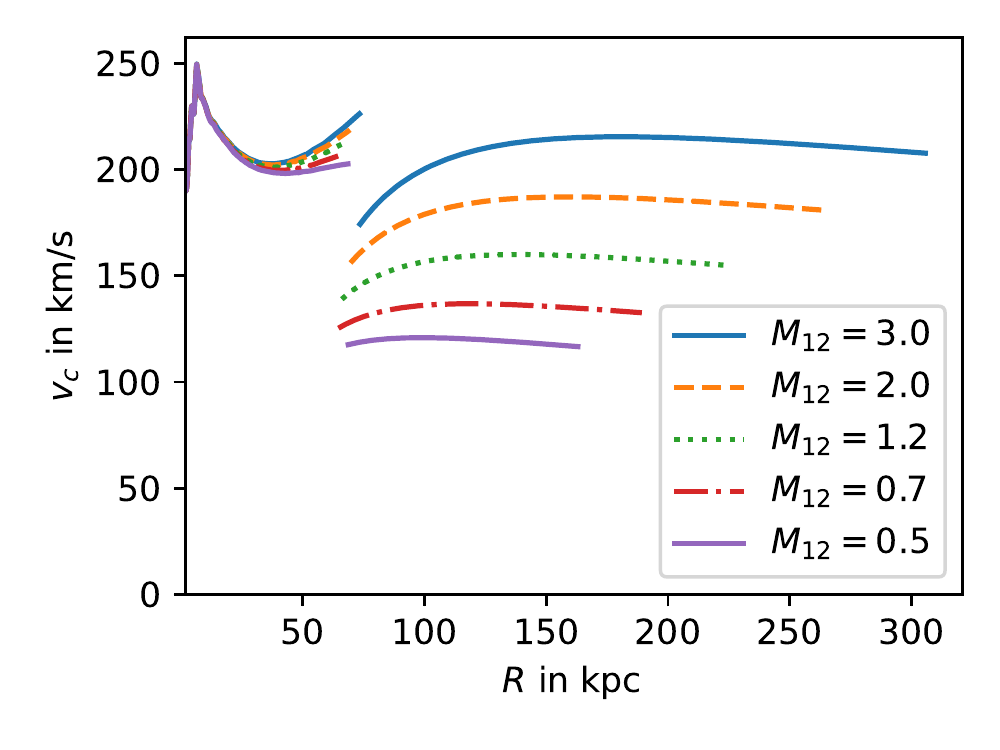}
 \caption{
        The rotation curves for different boundary conditions, corresponding to different total dark matter masses $M_{200}^{\rm{DM}} \equiv M_{12} \cdot 10^{12} \,M_\odot$.
        Each rotation curve is plotted for $R < r_{200}$ with the virial radius $r_{200}$.
        $ M_{200}^{\rm{DM}}$ and $r_{200}$ are calculated assuming the {\sc NFW} radius as the transition radius.
        The discontinuity at $R = R_{\rm{NFW}}$ is because we assume the phonon force is switched off at this radius.
        See also Table~\ref{tab:largeradii}.
 }
 \label{fig:SFDMlargeradii}
\end{figure}

\begin{table}
\center
\caption{
   The results of the calculation of the {\sc NFW} radius $R_{\rm{NFW}}$, the total dark matter mass $M_{200}^{\rm{DM}}$, the virial radius $r_{200}$, and the thermal radius $R_T$ for various boundary conditions $\mu_\infty$ imposed at a radius $r_\infty$.
    $ M_{200}^{\rm{DM}}$ and $r_{200}$ are calculated assuming the {\sc NFW} radius as the transition radius.
    The baryonic density is kept fixed with $f_b = 0.8$.}
 
\label{tab:largeradii}
\begin{tabular}{c|c|c|c|c|c|}
\hline
$r_\infty$ & $\mu_\infty/m$ & $R_{\rm{NFW}}$ & $M^{\rm{DM}}_{200}$ & $r_{200}$ & $R_T$ \\
$\rm{kpc}$ & $10^{-8}$ & $\rm{kpc}$ & $10^{12}\,M_\odot$ & $\rm{kpc}$ & $\rm{kpc}$ \\
\hline
110 & 7.80 & 73 & 3.0 & 306 & 105 \\
100 & 6.24 & 70 & 2.0 & 265 & 97 \\
100 & 1.25 & 66 & 1.2 & 225 & 87 \\
90 & 0.25 & 65 & 0.7 & 189 & 76 \\
80 & 0.12 & 69 & 0.5 & 163 & 67 \\
\hline
\end{tabular}
\end{table}

For $ f_b = 0.8 $, we found the thermal radius to be $ R_{\rm{T},R} = 87.5\,\rm{kpc} $ and $ R_{\rm{T}, z} = 87.6\,\rm{kpc} $.
Thus, the superfluid's thermal radius is almost the same in $R$- and $z$-direction.
This is not surprising since we assumed spherical symmetry at large radii.
Also, $ R_{\rm{T},R} $ is much larger than $ 25\,\rm{kpc} $.
This indicates that our above procedure for calculating the rotation curve is justified.

For $ f_b = 0.8 $, we found the {\sc NFW} radius to be $ R_{\rm{NFW},R} = 66.0\,\rm{kpc} $ and $ R_{\rm{NFW},z} = 65.7\,\rm{kpc} $.
Thus, the superfluid seems to be sufficiently spherically symmetric at the {\sc NFW} radius for our procedure to make sense.
Just as the thermal radius $ R_{\rm{T},R} $, the {\sc NFW} radius $ R_{\rm{NFW},R} $ is much larger than $ 25\,\rm{kpc} $ indicating that our above procedure for calculating the rotation curve is justified.

The difference between the {\sc NFW} and the thermal radii is about $ 30\,\% $, similar to the {\sc NFW} and thermal radii of the spherically symmetric galaxies studied in \citet{Berezhiani2018}.
Therefore, we should take these radii as rough estimates rather than precise values, as discussed in \cite{Berezhiani2015, Berezhiani2018}.
This is a limitation of this approach which assumes all of the dark matter particles to be in the condensed, superfluid phase in the inner parts of a galaxy with a sharp transition to the non-condensed, normal phase at larger radii.
It may be possible to improve on this using a two-component approach where one component is in the superfluid phase and one component is in the normal, not-condensed phase.
Nevertheless, the above suggests that the {\sc NFW} and thermal radii give a reasonable first approximation for the transition also in non-spherically-symmetric galaxies.

\subsection{Estimates for the total dark matter mass}
\label{sec:total}

So far, we used a fixed boundary condition $\mu_\infty/m = 1.25\cdot 10^{-8}$ at $r_\infty = 100\,\rm{kpc}$.
The reason is that the rotation curve at $R < 25\,\rm{kpc}$ depends only very weakly on this boundary condition.
However, the same is not true for the size of the superfluid core and the dark matter profile outside this superfluid core.
A similar observation was previously discussed in Secs.~6.2 and 6.3 of \cite{Hossenfelder2018}.
Therefore,  we here use several different boundary conditions to illustrate the behavior of {\sc SFDM} on larger scales.
For each choice of boundary condition, we estimate the thermal radius, the {\sc NFW} radius, the total dark matter mass, and the virial radius.
To keep a reasonable fit to the rotation curve at $ R < 25\,\rm{kpc}$, we fix $f_b = 0.8$.

Following \citet{Berezhiani2018}, we assume that the {\sc NFW} halo is matched to the superfluid core at the {\sc NFW} radius $R_{\rm{NFW}}$.
For concreteness, we take $R_{\rm{NFW}} \equiv R_{\rm{NFW},R}$.

To estimate the virial radius and the total mass of dark matter in both the superfluid and normal phase, we use the density profile given by $ \rho_{\rm{SF}} $ for $ r < R_{\rm{NFW}} $ and the {\sc NFW} profile $ \rho_{\rm{NFW}} $ matched at $ r = R_{\rm{NFW}} $ for $ r > R_{\rm{NFW}} $. With this, we can calculate the virial mass as
\begin{align}
 M^{\rm{DM}}_{200} = \, &2 \pi \iint\limits_{R^2+z^2 < R_{\rm{NFW}}^2} dR \, R \, dz \, \rho_{\rm{SF}}(R, z) \\ \nonumber
                     &+ 4 \pi \int\limits_{R_{\rm{NFW}}}^{r_{200}} dr \, r^2 \, \rho_{\rm{NFW}}(r) \,.
\end{align}
Here, $ r_{200} $ is the spherical radius where the mean dark matter density drops below $ 200 \cdot 3 H^2/(8 \pi G) $, where $H$ is the Hubble constant.
Here, we use $ H = 67.3\,\rm{km}/(\rm{s} \cdot \rm{Mpc}) $.

We choose a set of boundary conditions that covers a range in $M_{200}^{\rm{DM}}$ from $0.5\cdot10^{12}\,M_\odot$ to $3\cdot10^{12}\,M_\odot$.
This covers the range of measured values given in \citet{Bland-Hawthorn2016}\footnote{
    Although it should be kept in mind that these measurements may not apply directly in {\sc SFDM} if they assume standard {\sc CDM}.
}.
The precise boundary conditions and results are given in Table~\ref{tab:largeradii}.
For each boundary condition, we show the corresponding rotation curve up to the virial radius in Fig.~\ref{fig:SFDMlargeradii}.

The rotation curves have discontinuities at $R=R_{\rm{NFW}}$ because in the model of \citep{Berezhiani2015, Berezhiani2018} the phonon force is assumed to be effective only inside the superfluid core.
Therefore, we include this force only for $R<R_{\rm{NFW}}$.
In a real galaxy, of course, this transition should be gradual and not abrupt.
How exactly this transition happens requires further theoretical work that is beyond the scope of the present paper.
Thus, the rotation curves shown in Fig.~\ref{fig:SFDMlargeradii} should not be taken too seriously around the discontinuity at $R = R_{\rm{RNFW}}$.
Away from this discontinuity the rotation curves should represent the {\sc SFDM} expectation.

   As one sees from Fig.~\ref{fig:SFDMlargeradii}, the rotation curves agree with each other for small radii, but deviate at larger radii, where the difference in the boundary conditions becomes important.
    We also see that the rotation curves bend upwards before $R = R_{\rm{RNFW}}$.
    This is due to the combination of the phonon force, which gives flat rotation curves, and the cored superfluid dark matter profile which starts to contribute significantly to the rotation curve at these radii.
    This was already observed in \citet{Berezhiani2018}.
    This effect increases with increasing $M_{200}^{\rm{DM}}$ which may rule out larger values of $M_{200}^{\rm{DM}}$.

A glance at Table~\ref{tab:largeradii} makes clear that the thermal radius depends more strongly on the boundary conditions than the {\sc NFW} radius.
    The {\sc NFW} radius varies from $65\,\rm{kpc}$ for $M^{\rm{DM}}_{200} = 0.7\cdot10^{12}\,M_\odot$ to $73\,\rm{kpc}$ for $M^{\rm{DM}}_{200} = 3\cdot10^{12}\,M_\odot$, while the thermal radius varies from $67\,\rm{kpc}$ for $M^{\rm{DM}}_{200} = 0.5\cdot10^{12}\,M_\odot$ to $105\,\rm{kpc}$ for $M^{\rm{DM}}_{200} = 3\cdot10^{12}\,M_\odot$.
    Still, for the range of parameters considered here, the thermal radius and the {\sc NFW} radius agree with each other to roughly $30\,\%$ as in \citet{Berezhiani2018}.

    We find that the virial radius $r_{200}$ varies from $163\,\rm{kpc}$ for $M^{\rm{DM}}_{200} = 0.5\cdot10^{12}\,M_\odot$ to $306\,\rm{kpc}$ for $M^{\rm{DM}}_{200} = 3\cdot10^{12}\,M_\odot$.
    Assuming M31 to have a roughly similar virial radius, these values indicate that, today, there is no overlap of the halos of MW and M31, since the distance to M31 is about $770\,\rm{kpc}$ \citep{Karachentsev2004}.
    However, for even larger MW (or M31) masses, their halos may overlap so that dynamical friction may become important.

\section{Discussion}
\label{sec:disc}

In this present work, we are concerned with the rotation curve at $ R < 25 \,\rm{kpc} $. In this range, the rotation curve  depends only weakly on $ \mu_\infty $. It then suffices to confirm that our choice of $ \mu_\infty $ gives a superfluid core extending to $ R > 25\,\rm{kpc} $ as well as a reasonable rotation curve. As discussed in the previous section, this is indeed the case.
The precise choice of $ \mu_\infty $, however, becomes important for comparison to data at larger radii, especially beyond the superfluid core.
For example, we saw explicitly that the total dark matter mass and the virial radius depend strongly on this choice.

Knowing the transition radius and the behavior of {\sc SFDM} at larger radii is important to predict the behavior of tracers of the gravitational potential at these larger radii.
For example, as discussed in \citet{Berezhiani2018}, tidal stellar streams may exhibit peculiar features due to crossing the transition radius.
The present work can be a first step towards making predictions for such features.
Similarly, the transition radius determines which satellite galaxies and globular clusters are affected by a {\sc MOND}-like External Field Effect ({\sc EFE}) \citep{Milgrom1983a, Famaey2012}, since a {\sc MOND}-like {\sc EFE} applies only inside the superfluid core in {\sc SFDM} \citep{Berezhiani2018}.

It must further be mentioned that while we compared the model of {\sc SFDM} to rotation curve data, in the Milky-Way this is not the only available data.
For example, there may be additional constraints from vertical acceleration measurements as discussed in \cite{Lisanti2019,Lisanti2019b}. These constraints are a serious problem for {\sc MOND}. However, for {\sc SFDM} they are not necessarily problematic. Because the superfluid interacts with the baryons, it should strictly speaking also rotate, which quite plausibly affects the vertical gradient of the phonon-force. However, we do not presently have a theoretical framework to handle a rotating two-component fluid in a gravitational potential, so, unfortunately, we cannot address this interesting constraint here.

Above, we adjusted the parameter $ f_b $ to give a reasonable Milky Way rotation curve.
For a proper statistical analysis, both the parameters of the baryonic mass distribution and the parameters of the {\sc SFDM} model should be fitted.
There is little reason to doubt that it is possible to obtain a good fit to the Milky Way rotation curve in this way because the model has four free parameters, whereas we were able to get a reasonable rotation curve with only one free parameter. However, attempting to fit the {\sc SFDM} parameters to the MW rotation curve makes no sense in isolation because changing the parameters will affect the goodness-of-fit to other astrophysical data. To address this point, one would need to do a global fit to all available data to identify the best-fit parameters, but this is beyond the scope of this present work.

\section{Conclusions}
\label{sec:conc}

We have shown here that superfluid dark matter which mimics {\sc MOND} with a phonon-force has no trouble explaining the newest data for the Milky-Way rotation curve. Superfluid dark matter provides a  fit of the rotation curve that is similarly good as {\sc MOND}, provided that the total baryonic mass is $10-20\%$ less than the current estimates of the stellar mass. This amount of baryonic mass in the Milky Way is currently within measurement uncertainty. However, in the future, with better measurements of the Milky-Way's baryonic mass, the here presented result may enable us to tell apart superfluid dark matter from {\sc MOND}.

We have further demonstrated that the superfluid core's size in axisymmetric galaxies can be estimated with a similar procedure as in spherically symmetric galaxies and we have calculated the total dark matter masses and virial radii for various boundary conditions.
This can be an important first step to understand how satellite galaxies, globular clusters, and tidal stellar streams behave in {\sc SFDM}.

\section*{Acknowledgements}

We thank Stacy McGaugh for helpful discussion and for kindly providing the data used in our analysis. SH gratefully acknowledges support from the German Research Foundation ({\sc DFG}). We also wish to thank an anonymous referee for helpful suggestions for improvement.

\section*{Data availability}

The data underlying this article will be shared on reasonable request to the corresponding author.

\bibliographystyle{mnras}
\bibliography{sfdm-milky-way-rotation-curve-mnras.bib}

\appendix

\section{Numerical Method}
\label{sec:numerical}

To numerically solve the equations summarized in Section \ref{sec:theory}, we run Mathematica 12's \citep{Mathematica12} PDESolve in the region $ R^2 + z^2 < r_\infty^2 $, $ z > 0 $ with $ r_\infty = 100\,\rm{kpc} $, unless stated otherwise.
The value of $ r_\infty $ is chosen such that it is much larger than the size of the stellar and gas disks.
It is possible to solve the equations only in the region $ z > 0 $ because in our approximation the Milky Way is symmetric under $ z \to -z $.

The Mathematica function PDEsolve operates on a triangulation of the region described above.
For this, we impose the following maximum areas of the triangular cells:
For $ R < 20\,\rm{kpc} $ and $ z < 0.5\,\rm{kpc} $, the maximum area is $ (0.05\,\rm{kpc})^2 $.
For $ R < 40\,\rm{kpc} $ and $ z < 5\,\rm{kpc} $, it is $ (0.2\,\rm{kpc})^2 $.
For $ R < 40\,\rm{kpc} $ and $ z < 10\,\rm{kpc} $, it is $ (1\,\rm{kpc})^2 $.
For $ R < 40\,\rm{kpc} $ and $ z < 20\,\rm{kpc} $, it is $ (2\,\rm{kpc})^2 $.
Otherwise, the maximum area is $ (10\,\rm{kpc})^2 $.

In the term of the form $ \vec{\nabla} ( g \, \vec{\nabla} \theta ) $, $ g $ is not a smooth function for $ \theta, \hat{\mu} \to 0 $ due to the square roots in $ g $.
Unfortunately, this causes Mathematica's PDESolve to fail with the error ``NDSolve::femdpop: The FEMStiffnessElements operator failed.'' We work around this as follows.
In the case of the full {\sc SFDM} model, we rewrite the model, including the boundary conditions, in terms of $ \hat{\mu}_{\rm{temp}}(R,z) = \hat{\mu}(R,z) + \Delta \hat{\mu} $, where $ \Delta \hat{\mu} $ is a constant.
This suffices to make PDESolve solve the equations in terms of $ \hat{\mu}_{\rm{temp}} $.
The solution for $ \hat{\mu}_{\rm{temp}} $ then gives a solution for $ \hat{\mu} $ by subtracting $ \Delta \hat{\mu} $.
We verified that our results do not depend on the choice of $ \Delta \hat{\mu} $.
In the case of the idealized {\sc MOND} limit, we simply shift $ \sqrt{(\vec{\nabla} \bar{\theta})^2} \to \sqrt{C + (\vec{\nabla} \bar{\theta})^2} $ with a constant $ C \ll (\vec{\nabla} \bar{\theta})^2 $.
We verified that $ C \ll (\vec{\nabla} \bar{\theta})^2 $ for the obtained solutions and that the choice of $ C $ does not affect our results.

\bsp	%
\label{lastpage}
\end{document}